\documentclass[superscriptaddress,amsmath, nofootinbib]{revtex4}
\usepackage{amsfonts}
\usepackage{graphicx}
\usepackage[latin1]{inputenc}
\usepackage{amsmath}
\usepackage{amssymb}
\usepackage{euscript}
\begin{document}


\title{Dynamics of a relativistic discrete body: rigidity conditions, and covariant equations of motion.}

\author{Alexei A. Deriglazov }
\email{alexei.deriglazov@ufjf.br} \affiliation{Depto. de Matem\'atica, ICE, Universidade Federal de Juiz de Fora,
MG, Brazil} 

\author{}

\date{\today}

\begin{abstract}
Rigidity conditions for a body considered as a discrete system of relativistic particles are proposed. They by themselves do not yet determine an evolution of the system, and some second-order equations must be added to them. Poincar\'e-covariant equations of motion compatible with these rigidity conditions are proposed and discussed. The resulting theory has the expected six dynamical degrees of freedom, allowing for more general motions than in Born's theory. Therefore, treating a relativistic body as a discrete system of particles could be a promising alternative to the standard approach based on Born's rigidity conditions.
\end{abstract}

\maketitle 



\vspace{5mm}

\hspace{45mm}{\it Dedicated to the memory of Igor Viktorovich Tyutin}

\vspace{5mm}

\section{Introduction. }

In Galilean-covariant mechanics, the distance between particles turns out to be an observer-independent notion. This allows us to immediately define a rigid body as a discrete system of particles, the distances between which cannot change during their evolution. The advantage of considering a discrete system is that its rigidity conditions from the point of view of classical mechanics represent the holonomic (that is velocity independent) constraints. Holonomic mechanics is a well-developed and studied issue in classical mechanics. This made it possible to develop a systematic procedure for deriving the equations of motion starting from the Lagrangian of a rigid body \cite{AAD_RB,AAD_23}, without the need for additional postulates or intuitive assumptions that have to be made when using the standard formalism \cite{Lei_1965,Lev_2023,Yeh_22}. In configuration space, this procedure yields second-order equations for the rotation matrix\footnote{It is worth noting here that in the final equations for the independent degrees of freedom (rotation matrix), there is no difference between treating the non-relativistic body as a discrete or continuous object.}. In phase space, these equations turn into the first-order Euler-Poisson equations, written for the rotation matrix and angular velocity. Using this procedure, several inaccuracies made when deriving equations for some classical problems in rigid body dynamics were corrected \cite{AAD23_8,AAD_2023_9,AAD23_3}.

A natural next step is to see to what extent these methods of rigid body dynamics can be transferred to the case of the special theory of relativity.

When moving from Galilean mechanics to special relativity,  we lose the observer-independent notion of a distance. The basic invariant in special relativity is the four-interval between two events instead of a distance between two particles, and to determine the latter, we need to know their positions at the same moment in time. However, the simultaneity of events in special relativity is observer-dependent, so the notion of distance is as well.  In fact, each timelike (unit) vector ${\mathbb P}^\mu$,  ${\mathbb P}^2=-1$ of Minkowski space whose points (events) are labeled by four-columns $y^\mu$ defines its own concept of simultaneity and distance, according to the equation \cite{Synge_1956}:
\begin{eqnarray}\label{rrb1}
({\mathbb P}, ~y_1-y)=0.  
\end{eqnarray}
This equation specifies the foliation of Minkowski space into three-dimensional hyperplanes on which, from the point of view of some inertial observer, lie the events simultaneous with a given $y^\mu_1$. To see this, let us consider an inertial frame whose origin is described by a straight worldline in the direction ${\mathbb P}^\mu$. Then ${\mathbb P}^\mu=(1, 0, 0, 0)$ in this frame, and Eq. (\ref{rrb1}) reads $y^0_1=y^0$, that is, the 
events $y^\mu_1$ and $y^\mu$ turn out to be simultaneous in this frame, and the interval between them turns into the distance between the particles: $\triangle s^2=-(y_1^0-y^0)^2+({\bf y}_1-{\bf y})^2=({\bf y}_1-{\bf y})^2$.  In a frame with the vector ${\mathbb Q}^\mu$ not collinear with ${\mathbb P}^\mu$, these events are no longer simultaneous. In view of this, the body of a certain shape at rest in one inertial frame will be observed as having a different shape in the frames moving relative to it, see, for 
instance \cite{Rin_1961,Ber_1942,Red_2006}.  Moreover, if the body, say, rotates, and the distances between its points remain constant in the original frame, this is no longer the case for other frames. For instance, consider a rod rotating in the $(x, y)$\,-plane with one end at the origin of the coordinate system and another moving according to the 
law: ${\bf r}(t)=(l\cos\omega t, ~ l\sin\omega t ~)$, where $\omega=\mbox{const}$, and $l=\mbox{const}<c/\omega$. Then its length is $L(t)=l$ in this frame. The observer moving with the velocity $v$ in the direction of the $x$\,-axis will observe the rod of a variable length 
\begin{eqnarray}\label{rrb0}
L'(t' )=l\sqrt{1-\frac{v^2}{c^2}\cos^2[\omega\lambda(t')]}, 
\end{eqnarray} 
where $t'$ is its laboratory time, and $\lambda(t')$ is the solution to the algebraic 
equation $c^2\lambda-lv\cos\omega\lambda=c\sqrt{c^2-v^2}t'$.

Thus, the difficulty in attempting to formulate the dynamics of a body in special relativity arises already at the moment of defining the very concept of a rigid body in the Lorentz-covariant theory: observers must agree, in an observer-independent manner,  which particular system of relativistic particles they will all call a rigid body? In the present work, we test the following agreement \cite{Gar_1952,Syn_1952}.
Since a body cannot be rigid for all observers, the natural alternative to define it is to make an observer-independent agreement on the choice of a special {\it inertial} frame, in which the distances between particles should not change during their evolution. 
In practice, this means to write the Lorentz-covariant relations that distinguish this special frame from others. Each observer would then be able to study such a system in his own laboratory, that is, in a covariant way.

The first attempt to write the rigidity conditions for a relativistic body considered as {\it a continuous} object was undertaken in a seminal work of M. Born in 1909 \cite{Born_1909}. He proposed hydrodynamic-like equations specifying a vector field in Minkowski space, which is tangent to the worldlines of the body's particles at each point, see also \cite{Ros_1947, Sal_1954}. However, soon afterward, it was shown \cite{Her_1910,Noe_1910} that Born's rigidity conditions are too restrictive to describe all expected movements of a body: the motion of a Born rigid body is generally determined by the trajectory of just one of its points. This means that the body has lost 3 of its classical 6 degrees of freedom \cite{Born_1910,Sal_1954}. The only motions consistent with these conditions are either so-called plane motions (translations without rotation) or group motions (relativistic analogue of pure rotations) \cite{Sal_1954,Ros_1947}. Several possibilities were considered to relax Born's conditions, by requiring them only on the surface of the body \cite{Man_2009,n1,n7}. The general evolution equations for a congruence of 2-surface of observers on a curved background and in an  arbitrary frame  were studied in \cite{Clar_2007}. Further discussion of these problems can be found in the recent works \cite{New_1959,Eri_1982,2,3,4,5,6,7,8,9}. Due to the complexity of the problem already at the kinematic level, the task of constructing exact dynamic equations, as far as I know, has not even been posed to date. 

The purpose of the present work is to attempt to formulate less restrictive than Born's rigidity conditions. To this aim, we proceed in a slightly different way, by considering a body as a {\it discrete} system of relativistic particles. Our rigidity conditions represent a system of holonomic and nonholonomic constraints imposed on the particles' worldlines, providing rigidity for the entire body, not just its boundary. These constraints by themselves do not yet determine an evolution of the system, and some second-order equations must be added to them. Following the ideas of non-relativistic formalism \cite{AAD_RB,AAD_23}, we propose 
Poincar\'e-covariant equations compatible with our rigidity conditions. We show that the complete system of equations has a number of expected properties, including the right non-relativistic limit, and the right number (six) of dynamical degrees of freedom. 

The equations found may also be useful in discussing the relationship between different formalisms used to describe rotating bodies and spinning particles in classical and relativistic approaches \cite{Dmy_2026,Sun_2026,Liu_2026,Iso_2026,Kir_2026,Hus_2026,Rak_2026,Hwi_2026,Zha_2026}.

{\bf Notation.}  Capital letters of the Latin alphabet $N, P, A, B, \ldots$ or Greek letters from the beginning of the alphabet $\alpha, \beta, \ldots$ are used to label particles. Greek letters $\mu, \nu, \rho$ run through the values $0, 1, 2, 3$, and are used to label coordinates of four-dimensional Minkowski space with the metric $\eta_{\mu\nu}=diagonal(-1, +1, +1, +1)$. Latin letters $i, j, k, \ldots$, run through the values $1, 2, 3$, and used to label spatial coordinates. Three-vectors are denoted using the bold letters. For instance, the position vector of the particle $N$ 
is $y_N^\mu(\tau)=(y_N^0, ~{\bf y}_N)=(y_N^0, ~y_N^i)=(y_N^0, y_N^1,y_N^2, y_N^3)$. A dot over any quantity means the time derivative of that 
quantity: $\dot y_N^\mu=\frac{d y_N^\mu}{d\tau}$.
Summation over particles is always explicitly stated: $\sum_{N=1}^{n}m_Ny_N^\mu$. Repeated indices $\mu, \nu, \rho$ and $i, j, k$ are summed unless otherwise indicated. Notation for the scalar product:  $(y_N, ~y_ P)=\eta_{\mu\nu}y_N^\mu y_P^\nu$; ~ $\dot y_N^2=\eta_{\mu\nu}\dot y_N^\mu \dot y_N^\nu$; ~ $({\bf y}_N, ~{\bf y}_P)=y_N^i y_P^i$.

\section{Rigidity conditions for a discrete system of relativistic particles.}

Let us consider a system of $n\ge 4$ particles not all lying on the same plane, and in the absence of external forces. Their evolution will be described using worldlines of Minkowski space taken in the parametric form: $y^\mu_N(\tau)$, $N=1, 2, \ldots , n$. Then we can work with manifestly covariant equations, that is, with equations on which the Lorentz group is realized linearly: $y'^\mu_N(\tau)=\Lambda^\mu{}_\nu y^\nu_N(\tau)$. 

The system of particles we call a relativistic body if they obey the following conditions: \par

\noindent {\bf I.} Their evolution is governed by reparametrization-invariant and Poincar\'e-covariant dynamical equations, which imply the impossibility for the particles to exceed the speed of light, and should reproduce the dynamics of a non-relativistic rigid body in the limit $c\rightarrow\infty$. \par

\noindent {\bf II.}  In the theory, there is a timelike four-vector integral of motion, say, ${\mathbb P}^\mu(y_1(\tau), \ldots , y_n(\tau), \dot y_1(\tau), \ldots , \dot y_n(\tau))$, such that for any solution, the following conditions are satisfied:
\begin{eqnarray}\label{rrb2}
({\mathbb P}, ~y_N(\tau)-y_K(\tau))=0, \qquad \mbox{for any} ~  N, K, ~ \mbox{and} ~  \tau.  
\end{eqnarray} \par

\noindent {\bf III.} For any solution, the spacelike intervals among all particles do not depend on $\tau$
\begin{eqnarray}\label{rrb3}
(y_N(\tau)-y_K(\tau))^2=a_{NK}=\mbox{const}.
\end{eqnarray} 

This definition provides the properties we need. Let $y^\mu_N(\tau)$, $N=1, 2, \ldots , n$ be some solution to such a problem, and ${\mathbb P}^\mu$ be the constant vector corresponding to this solution. As we saw in the Introduction, there is an inertial observer for 
which ${\mathbb P}^\mu=({\mathbb P}^0, 0, 0, 0)$, ${\mathbb P}^0=$ const. We will call it a {\it comoving observer} of this solution. In this Laboratory the simultaneity constraints  ({\ref{rrb2}) imply 
\begin{eqnarray}\label{rrb4}
y_N^0(\tau)=y_K^0(\tau), \qquad \mbox{for any} ~  N, K, ~ \mbox{and} ~  \tau, 
\end{eqnarray} 
that is at each instant $\tau$ the events $y_N^\mu(\tau)$ and $y_K^\mu(\tau)$ are simultaneous. Then the rigidity constraints (\ref{rrb3}) state that distances among the particles do not change with time in this inertial frame
\begin{eqnarray}\label{rrb5}
({\bf y}_N(\tau)-{\bf y}_K(\tau))^2=a_{NK}=\mbox{const}, \qquad \mbox{for any} ~  N, K, ~ \mbox{and} ~  \tau.  
\end{eqnarray} 
Since the defining relations are Lorentz-invariant, our definition is given in an observer-independent manner. Therefore, the equations (\ref{rrb2}) and (\ref{rrb3}) can be adopted by all observers as the rigidity conditions defining a relativistic rigid body. 
%
%

\section{Deduction of covariant dynamical equations of a relativistic body.}

Here, we look for an example of a dynamical theory satisfying our rigidity conditions. 
We start from a non-relativistic rigid body that can be described by the Lagrangian action \cite{AAD_RB,AAD_23}
\begin{eqnarray}\label{rrb6}
S_{nr}=\int dt ~\frac12\sum_{N=1}^{n}m_N\dot{\bf y}_N^2+
\frac12\sum_{\alpha=2}^{4}\sum_{B=2}^{n}\lambda_{\alpha B}\left[({\bf y}_\alpha-{\bf y}_1, ~{\bf y}_B-{\bf y}_1)-a_{\alpha B}\right]. 
\end{eqnarray} 
The first term is the kinetic energy of all particles, while the second term is composed of holonomic constraints added to the action using the Lagrangian multipliers $\lambda_{\alpha B}(t)$. The constraints are constructed with the use of four particles ${\bf y}_1, {\bf y}_2, {\bf y}_3$,  and ${\bf y}_4$, selected among ${\bf y}_N$, and not lying on the same plane. The constraints are functionally independent and guarantee that the distances among all pairs of particles do not depend on time. Besides, they imply that the whole system has six independent degrees of freedom \cite{AAD_RB,AAD_23}. 

To move to a relativistic theory, we first replace the kinetic terms with their relativistic counterparts as follows:
\begin{eqnarray}\label{rrb7}
S_3=\int dt ~ -\sum_{N=1}^{n}m_Nc\sqrt{c^2-\dot{\bf y}_N^2}+
\frac12\sum_{\alpha=2}^{4}\sum_{B=2}^{n}\lambda_{\alpha B}\left[({\bf y}_\alpha-{\bf y}_1, ~{\bf y}_B-{\bf y}_1)-a_{\alpha B}\right]. 
\end{eqnarray} 
For the latter use, we present an explicit form of dynamical equations implied by this action
\begin{eqnarray}\label{rrb8}
m_Nc\frac{d}{dt}\left(\frac{\dot{\bf y}_N}{\sqrt{c^2-\dot{\bf y}_N^2}}\right)={\bf F}_N, 
\end{eqnarray}
where 
\begin{eqnarray}\label{rrb9}
{\bf F}_1=-\sum_{\alpha, \beta=2}^{4}\lambda_{\alpha\beta}[{\bf y}_\beta-{\bf y}_1]-
\frac12\sum_{\beta=2}^{4}\sum_{\alpha=5}^{n}\lambda_{\beta\alpha}[{\bf y}_\beta+{\bf y}_\alpha-2{\bf y}_1]; \cr
{\bf F}_\alpha=\sum_{\beta=2}^{4}\lambda_{\alpha\beta}[{\bf y}_\beta-{\bf y}_1]+
\frac12\sum_{\beta=5}^{n}\lambda_{\alpha\beta}[{\bf y}_\beta-{\bf y}_1],  \quad \alpha=2, 3, 4; \cr
{\bf F}_\beta=\frac12\sum_{\gamma=2}^{4}\lambda_{\beta\gamma}[{\bf y}_\gamma-{\bf y}_1], \quad \beta=5, 6, \ldots , n.
\end{eqnarray}
The forces obey the identity $\sum_{N=1}^{n} {\bf F}_N=0$,  which together with the constraints implies $\sum_{N=1}^{n} \dot y_N^iF_N^i=0$.
 In turn, these relations imply the conservation of a total linear momentum, energy, and angular momentum of the system.

By construction, particles of this body can not exceed the speed of light in the process of their movement. However, in the three-dimensional formulation, it is not clear whether this theory is Lorentz-covariant (and as we will see, without additional assumptions, it is not). It seems instructive to discuss this point in more detail.  Let us proceed similarly to the case of a unique relativistic particle \cite{GT,deriglazov2010classical}, simply replacing all three-dimensional quantities with their four-dimensional counterparts: ${\bf y}_N(t)\rightarrow y^\mu_N(\tau)$. Then (\ref{rrb7}) turns into the manifestly covariant action 
\begin{eqnarray}\label{rrb12}
S_4=\int d\tau ~ -\sum_{N=1}^{n}m_Nc\sqrt{-\dot  y_N^2}+
\frac12\sum_{\alpha=2}^{4}\sum_{B=2}^{n}\lambda_{\alpha B}\left[( y_\alpha- y_1, ~ y_B- y_1)-a_{\alpha B}\right],  
\end{eqnarray}
which implies the following dynamical equations: 
\begin{eqnarray}\label{rrb13}
m_Nc\frac{d}{d\tau}\left(\frac{\dot y^\mu_N}{\sqrt{-\dot y_N^2}}\right)= F^\mu_N. 
\end{eqnarray}
Unfortunately, this manifestly covariant theory is not suitable for our purposes. Indeed, we are interested in equations for which the conditions (\ref{rrb4}) are satisfied in some inertial frame. Assuming that our covariant theory is considered in this frame, let us denote the resulting unique null-coordinate as $y^0(\tau)$. Using (\ref{rrb4}) in equations (\ref{rrb13}) we get $F^0_N=0$, and then in addition to $3n$ expected equations (\ref{rrb8}), in this theory will be presented $n$ extra equations 
$d\left(\dot y^0/\sqrt{(\dot y^0)^2-\dot {\bf y}_N^2}\right)/d\tau=0$.  Returning to the physical-time parametrization $y^0=ct$, $\tau=t$, 
we get $d\left(1/\sqrt{c^2-\dot {\bf y}_N^2}\right)/dt=0$, $N=1, 2, \ldots , n$. This implies 
$\dot{\bf y}_N^2=\mbox{const}$ for each $N$, that is, the three-dimensional speeds of the body's particles do not change during their evolution. Therefore, such a theory will be able to describe only the simplest movements, like pure translations or rotations. This resembles the situation with Born's conditions, which only allow either planar or group motions (see also \cite{Mat_1953}).

The above discussion gives a hint at how the desired theory might be obtained: the equations (\ref{rrb13}) should be modified in such a way that 
their null-components do not lead to $n$ additional equations on physical variables ${\bf y}_N$. Taking into account the reparametrization invariance, we need $n-1$ auxiliary variables that enter essentially only into the null-components of the equations (\ref{rrb13}), and allow us to satisfy them with the help of auxiliary variables instead of ${\bf y}_N$.

Let us look for such equations. First, as independent simultaneity constraints  (\ref{rrb2}) we take $n-1$ relations
\begin{eqnarray}\label{rrb14}
({\mathbb P}, ~y_B-y_1)=0, \qquad B=2, 3, \ldots , n,  
\end{eqnarray}
where for ${\mathbb P}^\mu$ we tentatively take the total linear momentum of a system of free relativistic particles 
\begin{eqnarray}\label{rrb15}
{\mathbb P}^\mu\equiv\sum_{N=1}^{n}\frac{m_Nc\dot y^\mu_N}{\sqrt{-\dot y_N^2}}. 
\end{eqnarray}
Second, we take the following scalar products as independent rigidity constraints: 
\begin{eqnarray}\label{rrb16}
(y_\alpha-y_1, ~y_B-y_1)=a_{\alpha B}=\mbox{const}, \qquad \alpha=2, 3, 4, \quad  B=2, 3, \ldots , n.
\end{eqnarray} 
Third, in addition to these first-order and algebraic equations, we postulate the following dynamical equations:
\begin{eqnarray}\label{rrb17}
m_1c\frac{d}{d\tau}\left(\frac{\dot y^\mu_1}{\sqrt{-\dot y_1^2}}\right)+{\mathbb P}^\mu(\sum_{B=2}^{n}\Lambda_B)= F^\mu_1, \qquad \qquad \qquad \label{rrb17.1} \\ 
m_Ac\frac{d}{d\tau}\left(\frac{\dot y^\mu_A}{\sqrt{-\dot y_A^2}}\right)+{\mathbb P}^\mu\Lambda_A= F^\mu_A,  \qquad A=2, 3, \ldots , n,  \label{rrb17.2} 
\end{eqnarray}
written for the original dynamical variables $y^\mu_N(\tau)$, and $n-1$ auxiliary scalar functions $\Lambda_A(\tau)$. The four-force $F^\mu_N$ appeared in this equations is obtained from (\ref{rrb9}) by replacing the three-vectors ${\bf y}_N(t)$ on $y^\mu_N(\tau)$. 
Using the notation $\Lambda_1\equiv-\sum_{B=2}^{n}\Lambda_B$, or, equivalently, $\sum_{N=1}^{n}\Lambda_N=0$, 
the equations (\ref{rrb17.1}) and (\ref{rrb17.2}) can be rewritten in a more symmetric form as follows: 
\begin{eqnarray}\label{rrb19}
m_Nc\frac{d}{d\tau}\left(\frac{\dot y^\mu_N}{\sqrt{-\dot y_N^2}}\right)+{\mathbb P}^\mu\Lambda_N= F^\mu_N,  \qquad 
N=1, 2, \ldots , n, \label{rrb19.1} \\
\sum_{N=1}^{n}\Lambda_N=0, \qquad \qquad \qquad\qquad \qquad\qquad \qquad \qquad\qquad  ~   \label{rrb19.2}
\end{eqnarray}
where $\Lambda_N(\tau)$ are $n$ scalar functions obeying the constraint (\ref{rrb19.2}). 

To obtain the final form of our equations, we eliminate these auxiliary dynamic variables from the system (\ref{rrb19.1}) and (\ref{rrb19.2}). Let us denote the relativistic momentum of the particle $y_N^\mu$ as follows: 
\begin{eqnarray}\label{rrb20}
p_N^\mu\equiv\frac{m_Nc\dot y^\mu_N}{\sqrt{-\dot y_N^2}}.   
\end{eqnarray}
Contracting (\ref{rrb19.1}) with ${\mathbb P}_\mu$ we obtain 
\begin{eqnarray}\label{rrb21}
\Lambda_N=\frac{{\mathbb P}_\mu}{{\mathbb P}^2}[\dot p_N^\mu-F_N^\mu], \qquad \mbox{then due to the identity} \quad 
\sum_{N=1}^{n}F_N^\mu=0 \quad \mbox{we get} \quad 
\sum_{N=1}^{n}\Lambda_N=\frac{{\mathbb P}_\mu\dot{\mathbb P}^\mu}{{\mathbb P}^2}. 
\end{eqnarray}
Substituting $\Lambda_N$ back into (\ref{rrb19.1}) and (\ref{rrb19.2}) we exclude the auxiliary variables and get  
\begin{eqnarray}\label{rrb22.1}
N^\mu{}_\nu({\mathbb P})[\dot p_N^\nu-F_N^\nu]=0,    \qquad 
{\mathbb P}_\mu\dot{\mathbb P}^\mu=0. 
\end{eqnarray}
Here appeared the projector on the hyperplane orthogonal to ${\mathbb P}^\mu$
\begin{eqnarray}\label{rrb23}
N^\mu{}_\nu({\mathbb P})\equiv\delta^\mu{}_\nu-\frac{{\mathbb P}^\mu{\mathbb P}_\nu}{{\mathbb P}^2}, \qquad \mbox{this implies} \quad 
N^\mu{}_\rho({\mathbb P})N^\rho{}_\nu({\mathbb P})\equiv N^\mu{}_\nu({\mathbb P}), \qquad 
N^\mu{}_\nu({\mathbb P}){\mathbb P}^\nu\equiv 0. 
\end{eqnarray}
Together with the projector $K^\mu{}_\nu\equiv {\mathbb P}^\mu {\mathbb P}_\nu/{\mathbb P}^2$ onto the direction of ${\mathbb P}^\mu$, they form a decomposition of the unit 
\begin{eqnarray}\label{rrb23.1}
\delta^\mu{}_\nu=N^\mu{}_\nu+K^\mu{}_\nu. 
\end{eqnarray}

To summarize, we propose to postulate the following equations describing a relativistic body in the special theory of relativity:
\begin{eqnarray}
N^\mu{}_\nu({\mathbb P})\left[\frac{d}{d\tau}\left(\frac{m_Nc\dot y^\nu_N}{\sqrt{-\dot y_N^2}}\right)-F_N^\nu\right]=0,  \qquad N=1, 2, \ldots , n, \qquad \qquad \quad \label{rrb24.1} \\
{\mathbb P}_\mu\dot{\mathbb P}^\mu=0,  \qquad \qquad \qquad \qquad \qquad \qquad \qquad  \qquad \qquad \qquad \qquad \qquad \quad ~ \label{rrb24.2} \\
 ({\mathbb P}, ~y_B-y_1)=0, \qquad B=2, 3, \ldots , n,  \qquad \qquad \qquad \qquad \qquad \qquad \quad ~ \label{rrb24.3} \\
(y_\alpha-y_1, ~y_B-y_1)=a_{\alpha B}=\mbox{const}, \qquad \alpha=2, 3, 4, \quad  B=2, 3, \ldots , n. ~\label{rrb24.4}
\end{eqnarray}
In these equations, ${\mathbb P}^\mu$ is the combination of particle's velocities (\ref{rrb15}), while the four-force $F^\nu_N$ is obtained from (\ref{rrb9}) by replacing the three-vectors ${\bf y}_N(t)$ on $y^\nu_N(\tau)$.

\section{Self-consistency checks.}

The proposed equations represent a mixed system composed of $4n+1$ second-order differential equations, $n-1$ first-order differential equations, and $3(n-1)$ algebraic equations for $4n$ functions $y_N^\mu(\tau)$. Besides, the higher derivatives are not separated into left-hand sides in these equations; that is, they do not have the normal form. For such a kind system, even the existence of solutions is not guaranteed in any way \cite{GT}. So, here we discuss the structure of these equations and perform several tests of their self-consistency, including the existence of solutions. 

The theory (\ref{rrb24.1})-(\ref{rrb24.4}) is manifestly Poincar\'e-covariant. Besides, our equations preserve their form under arbitrary change of the evolution parameter: $\tau\rightarrow \tau'=\varphi(\tau)$, that is, they are reparametrization-invariant. 

Due to the projector (\ref{rrb23}) presented in $4n$ equations of second order (\ref{rrb24.1}), they obey $n$ 
identities: ${\mathbb P}_\mu N^\mu{}_\nu({\mathbb P})\left[\dot p_N^\nu-F_N^\nu\right]\equiv 0$, $N=1, 2, \ldots , n$. Therefore, among them, 
only $3n$ equations are independent. The second-order equation (\ref{rrb24.2}) implies that ${\mathbb P}^2(\dot y^\mu_N)$ is a scalar integral of motion of the system. The first-order equations (\ref{rrb24.3}) together with the algebraic equations (\ref{rrb24.4}) imply that the whole system has six independent degrees of freedom, as it should be for a rigid body. 

{\bf Four-vector integral of motion.} Summing up the equations (\ref{rrb24.1}) and taking into account 
that $\sum_{N} p^\mu_N={\mathbb P}^\mu$ and $\sum_{N} F^\mu_N\equiv 0$, we get the consequence $N^\mu{}_\nu\dot{\mathbb P}^\nu=0$. 
Eq. (\ref{rrb24.2}) can be equivalently rewritten in the form $K^\mu{}_\nu\dot{\mathbb P}^\nu=0$. Taking into account (\ref{rrb23.1}), we conclude that $\dot {\mathbb P}^\nu=0$, so ${\mathbb P}^\nu=\mbox{const}$. Therefore, the quantity (\ref{rrb15}) is a four-vector integral of motion of our equations. Being a sum of timelike vectors, it is also timelike. 

Taking into account that equations (\ref{rrb24.3}) and (\ref{rrb24.4}) imply (\ref{rrb2}) and (\ref{rrb3}), we conclude that our theory satisfies all desired rigidity conditions.

{\bf Equations in terms of laboratory time in the comoving frame.}  
To confirm the compatibility of the equations (\ref{rrb24.1})-(\ref{rrb24.4}), we discuss the search for a solution for which the conserved total 
momentum (\ref{rrb15}) is of the form: 
\begin{eqnarray}\label{rrb25}
{\mathbb P}^\mu=({\mathbb P}^0, 0, 0, 0).
\end{eqnarray}
According to the previous discussions, such a solution will describe a  
movement of the body that appears rigid in this very laboratory.
Since the total momentum is conserved in our theory, it is sufficient to satisfy the equality (\ref{rrb25}) at the initial moment of time. That is, we are looking for a solution with the initial data for $\dot y^\mu_N(0)=v_N^\mu$ satisfying the following conditions:
\begin{eqnarray}\label{rrb25.1}
\sum_{N=1}^{n}\frac{m_Nc v^0_N}{\sqrt{-v_N^2}}={\mathbb P}^0,  \qquad 
\sum_{N=1}^{n}\frac{m_Ncv^i_N}{\sqrt{-v_N^2}}=0.  
\end{eqnarray}
With ${\mathbb P}^\mu$ of the special form (\ref{rrb25}), Eq. (\ref{rrb24.3}) implies $y^0_B(\tau)=y^0_1(\tau)$, $B=2, 3, \ldots , n$. Let us denote the resulting unique null-coordinate as $y^0(\tau)$. Further, since the theory is reparametrization invariant, we choose the Laboratory time as a parameter: 
\begin{eqnarray}\label{rrb26}
y^0=ct, \qquad \tau=t. 
\end{eqnarray}
Substituting (\ref{rrb25}) and (\ref{rrb26}) into (\ref{rrb24.1}), (\ref{rrb24.4}), and (\ref{rrb24.2}), we get the following (nonvanishing identically) equations: 
\begin{eqnarray}\label{rrb27}
m_Nc\frac{d}{dt}\left(\frac{\dot{\bf y}_N}{\sqrt{c^2-\dot{\bf y}_N^2}}\right)={\bf F}_N,  \qquad \mbox{or, equivalently,} \quad 
m_Nc\frac{d^2y_N^i}{dt^2}=\sqrt{c^2-\dot{\bf y}_N^2}\left(\delta^i{}_j-\frac{\dot y_N^i\dot y_N^j}{c^2}\right)F_N^j. 
\end{eqnarray}
\begin{eqnarray}\label{rrb28}
({\bf y}_\alpha-{\bf y}_1, ~{\bf y}_B-{\bf y}_1)=a_{\alpha B}=\mbox{const}, \qquad \alpha=2, 3, 4, \quad  B=2, 3, \ldots , n. 
\end{eqnarray}
\begin{eqnarray}\label{rrb29}
\frac{d}{dt}\left(\sum_{N=1}^{n}\frac{m_Nc^2}{\sqrt{c^2-\dot{\bf y}_N^2}}\right)=0. 
\end{eqnarray}
The equation (\ref{rrb24.2}) in the comoving frame has become the condition (\ref{rrb29}) for conservation of total relativistic energy. But this is a consequence of the equations (\ref{rrb27}), and therefore can be omitted (indeed, Eqs.  (\ref{rrb27}) imply  (\ref{rrb29}) due to the 
identity $d(c^2/\sqrt{c^2-\dot{\bf y}^2})/dt=\dot{\bf y}d(\dot{\bf y}/\sqrt{c^2-\dot{\bf y}^2}~)/dt$~). 

As a result, our covariant equations in the comoving frame and in the physical-time parameterization are reduced to (\ref{rrb27}) and (\ref{rrb28}). Given a number ${\mathbb P}^0$, they should be solved with initial data for velocities satisfying the following conditions: 
\begin{eqnarray}\label{rrb30}
\sum_{N=1}^{n}\frac{m_Nc^2}{\sqrt{c^2-{\bf v}_N^2}}={\mathbb P}^0,  \qquad
\sum_{N=1}^{n}\frac{m_Nc{\bf v}_N}{\sqrt{c^2-{\bf v}_N^2}}=0.
\end{eqnarray}
{\it Solutions to the system (\ref{rrb27}) and (\ref{rrb28}) with other initial conditions do not describe any motion of our relativistic body.}

Note that the equations (\ref{rrb27}) and (\ref{rrb28}) follow from the variational problem (\ref{rrb7}). The theory has the expected non-relativistic limit: the equations (\ref{rrb27}) and (\ref{rrb28}) in the limit $c\rightarrow\infty$ turn into the standard equations of a non-relativistic rigid body (the latter are obtained from the Lagrangian action (\ref{rrb6}), see \cite{AAD_RB,AAD_23}). 

The equations (\ref{rrb27}) represent a normal system of $3n$ equations for $3n$ variables. The right-hand sides of these equations take into account the presence of holonomic constraints (\ref{rrb28}). The compatibility of such equations and the existence of solutions are well-established facts of classical mechanics \cite{Arn_1,Rub_1957, deriglazov2010classical}. 

In the comoving frame, some motions in the theory possess properties almost identical to those of the non-relativistic case. These are the motions in which the speed of each particle remains constant throughout the motion, so the corresponding Lorentz-contraction factor is as well. The equations in the comoving inertial frame reduce to (\ref{rrb8}). With the constant Lorentz-contraction factors, they simply coincide with the non-relativistic equations in which particle masses are replaced by relativistic masses. In particular, there are possible stationary rotations with angular 
velocity $\omega$ of a homogeneous ball of radius $r<c/\omega$. The latter restriction is due to the Lorentz-contraction factors appearing 
in Eqs. (\ref{rrb8}).  

{\bf Dynamics of the relativistic body in terms of irreducible degrees of freedom}. Of course, $3n$ equations like (\ref{rrb27}) for six independent degrees of freedom are too complicated for practical calculations and analysis. We need a set of more convenient variables. To find them, let us select some point rigidly connected to the body, say, ${\bf y}_c(t)$. Due to the relations (\ref{rrb28}), any solution to the system (\ref{rrb27})  and (\ref{rrb28}) is of the 
form ${\bf y}_N(t)={\bf y}_c(t)+R(t){\bf x}_N(0)$, where $R(t)$ is an orthogonal matrix, while ${\bf x}_N(0)={\bf y}_N(0)-{\bf y}_c(0)$ are the initial positions of the body's particles relative to the selected point ${\bf y}_c(0)$. Substituting these expressions for ${\bf y}_N(t)$ into Eqs. (\ref{rrb27}), we obtain a set of equations involving new variables ${\bf y}_c(t)$,  $R(t)$, and Lagrangian multipliers $\lambda_{\alpha B}(t)$. Further, following the procedure described in \cite{AAD_RB,AAD_23}, we can eliminate all auxiliary variables $\lambda_{\alpha B}(t)$, arriving at closed equations of motion for the physical variables ${\bf y}_c(t)$ and  $R(t)$. We hope to do this in the future.

{\bf Behavior of the relativistic body in an arbitrary inertial frame.} Let us discuss what the motion of our body looks like for an arbitrary inertial observer, who will study the equations (\ref{rrb24.1})-(\ref{rrb24.4}). Let $y^\mu_N(\tau)$, $N=1, 2, \ldots , n$ be a solution to these equations, and ${\mathbb P}^\mu(\dot y_N^\nu(\tau))=\mbox{const}$ be the corresponding conserved momentum. There are two possibilities. \par
\noindent A.  Let ${\mathbb P}^\mu=({\mathbb P}^0, 0, 0, 0)$. Then the body will look rigid in this inertial frame. \par 
\noindent B. Let ${\mathbb P}^\mu$ have no special form specified above. For further discussion, it is convenient to select an independent parameter on each world line: $y^\mu_N(\tau_N)$. Using reparametrization invariance, the observer will choose the laboratory time as a parameter on one of the worldlines, say, $N=1$: $\tau_1\rightarrow ct=y^0_1(\tau_1)$. Now it is written like this: $(ct, ~ y_1^i(t) )$. 
For each of the remaining world lines, the observer will resolve the equation $y^0_B(\tau_B)=ct$ with respect to $\tau_B$: $\tau_B=\tilde y^0_B(t)$, and substitute $\tau_B$ back into the functions $y^\mu_B(\tau_B)$. On all world lines, the events with the same value of $t$ are now simultaneous in this 
Laboratory: $(ct, ~ y^i_N(\tilde y^0_N(t))$, $N=1, 2, \ldots , n$. Having calculated the scalar products 
\begin{eqnarray}\label{rrb31}
(  {\bf y}_N(t)-{\bf y}_K(t), ~{\bf y}_L(t)-{\bf y}_M(t) ),
\end{eqnarray}
our observer will know how the distances and angles between particles of the relativistic body change during its motion in his Laboratory. 

To summarize, for a system of relativistic particles, we proposed the rigidity conditions (\ref{rrb2}) and (\ref{rrb3}), allowing us to follow the ideas of non-relativistic formalism when deriving the equations of motion (\ref{rrb24.1})-(\ref{rrb24.4}). 
We have shown that treating a relativistic body as a discrete system of particles could be a promising alternative to the standard approach based on Born's rigidity conditions.

\end{document}